# Characterization of Two-photon Photopolymerization Fabrication using High-speed Optical Diffraction Tomography

Yanping He, [a, 1] Qi Shao, [b, 1] Shih-chi Chen, [b, *] Renjie Zhou [a,*]

[a] Department of Biomedical Engineering, The Chinese University of Hong Kong, Shatin, New Territories, Hong Kong SAR, China
[b] Department of Mechanical and Automation Engineering, The Chinese University of Hong Kong, Shatin, New Territories, Hong Kong, China
[1] These authors contributed equally to this work
[*] Corresponding author. Email: scchen@mae.cuhk.edu.hk, rjzhou@cuhk.edu.hk
Postal address: Room 1111, William M.W. Mong Engineering Building, The Chinese University of Hong Kong, Shatin, N.T., Hong Kong

## Abstract

Two-photon photopolymerization (TPP) has recently become a popular method for the fabrication of three-dimensional (3D) micro- and nanostructures. The reproduction fidelity of the designed micro- and nanostructures is influenced by experimental writing conditions, including laser power, exposure time, etc. To determine the appropriate writing parameters, characterization of morphological features and surface roughness during the experiment is needed. Traditional characterization methods for TPP, e.g., scanning electron microscopy and atomic force microscopy, have limited speed and cannot study internal structures without invasive approaches. Optical diffraction tomography (ODT) is an emerging label-free 3D imaging technique based on reconstructing the object's 3D refractive index (RI) distribution with diffraction-limited resolution. Here, we propose a non-invasive solution to fully characterize the TPP-fabricated structures using a high-speed ODT technique, which can eliminate the need for complex sample preparation, such as fluorescence labelling or metal-coating, and achieve a full 3D measurement time of 6 ms. By visualizing and studying different TPP-fabricated structures, including embedded spirals and cubes, via the ODT system, the fabrication quality, including 3D morphological features, exposure levels, and surface roughness, can be examined quantitatively. The results suggest our method can effectively improve the fabrication quality and reproducibility of TPP, generating impacts on the nanofabrication community.

## Keywords

3D printing; two-photon polymerization; high-speed characterization; optical diffraction tomography

## 1. Introduction

As a direct laser writing technique, two-photon photopolymerization (TPP) has become one of the most widely used manufacturing techniques for the fabrication of three-dimensional (3D) micro- and nanostructures with sub-micron resolution [1, 2] in recent years. During the TPP process, the polymerization is initiated on a photosensitive resin by the nonlinear two-photon absorption effect that only occurs at the sub-micron focal region of a femtosecond laser [1]. With its excellent focal confinement, TPP has been broadly applied to creating complex microscale or nanoscale devices in 3D for use in different fields, such as microfluidic channels [3], meta-structures [4], and photonic crystals [5]. Recently, a new high-speed TPP method based on a digital micromirror device (DMD) that enables random-access scanning has been developed by our team [6, 7]. By utilizing a series of binary holograms (i.e., Lee holograms) [8], one to tens of scanning laser foci can be generated in the resin at the DMD refresh rate at over 20 kHz. Using this TPP method, we demonstrated high-speed parallel nanofabrication [6] and aberration-free large-area stitch-free 3D nano-printing [7]. As the properties and functions of a fabricated structure are often influenced by its morphology and refractive index (RI) distribution, characterization is needed to precisely examine the fabricated structure and ensure the result is consistent with the design.

To date, a few methods have been used to characterize TPP-printed structures by offering different resolving capabilities, including scanning electron microscopy (SEM) [9], atomic force microscopy (AFM) [10], focused ion beam sectioning (FIB) [11], x-ray computed tomography (CT) [12], prism coupling method [13], and laser interferometry [14]. However, these methods have various limitations, e.g., low measurement throughput, high sample invasiveness, sample contact, and small field of view. More importantly, most of the methods cannot obtain the internal features, but only the surface shapes. For example, SEM is frequently used for mapping the surface shapes with a high resolution, but the sample often needs to be metal-coated to increase conductivity, which damages the sample. AFM can also obtain the surface profiles of TPP structures with less invasiveness, but its serial detection process is time-consuming when mapping a small volume, e.g., ~1.5 hours to scan a volume of ~150 ×150 × 20 $\mu m^3$ using AFM (NanoWizard NanoOptics, Bruker). On the other hand, for internal feature detection, FIB is often used. Although effective, this method is invasive and of low yield and high cost. Recently, x-ray CT has been used to characterize the internal structures of a printed part [12]. However, most photoresins used in TPP have no contrasts for x-rays and custom-developed radiopaque resins are required. To obtain the RI information, an extra optical system is often built based on the prism coupling method [13] or optical interferometry method [14], while specific structures are printed for obtaining the RI measurements (e.g., thick layers, prism, etc.). Therefore, a fast 3D characterization method that can simultaneously obtain surface shape and roughness, internal features, and RI distribution for arbitrarily-shaped TPP-printed structures is yet to be developed.

Optical diffraction tomography (ODT) has emerged as a powerful tool for label-free 3D characterization of biological specimens [15, 16]. In ODT, the 3D RI distribution of a specimen is reconstructed by measuring the scattered fields coming at different illumination angles [17, 18], sample rotations [19], or focal depths [20-22]. Illumination angle scanning based ODT can achieve much higher volumetric imaging speed without perturbing the sample when compared with sample rotation or focus scanning [15]. In recent years, by displaying binary holograms on DMDs, stable and fast illumination angle scanning has been achieved in ODT [23, 24]. When matched with a high-speed camera, the angle scanning and acquisition rate of interferograms can achieve over 10 kHz, thus achieving volumetric imaging at >200 volumes per second (vps) [25], or > 600 vps when angle multiplexing is applied [26], or even >10,000 vps through machine learning [27].

In this work, we present a new method to characterize TPP printed structures based on a high-speed ODT system (first reported in a conference paper [25]). The ODT system is integrated with a custom-built DMD-based TPP fabrication system [6, 7] that achieves multi-focus random-access scanning. To evaluate the performance of the ODT-TPP system, we have designed several micro-structures with embedded complex 3D structures via different photoresins (of different RI values). The printed structures are visualized by the high-speed ODT module. The results show that complete 3D morphological information, including internal structures and surface geometry and roughness, of all fabricated structures can be retrieved from the measured 3D RI distributions. For comparison and validation, we used an SEM to map the structure surfaces and an AFM to quantify the surface topography. The measurements from ODT module are consistent with results from the SEM and AFM. Notably, the ODT module performed the measurements in only 6 ms and with minimal sample preparation and provided internal structure and 3D RI distribution information that cannot be realized by the SEM and AFM.

## 2. System design

Figure 1 presents the design of the ODT-TPP system. The laser source for TPP is an 80 MHz Ti: Sapphire femtosecond (fs) laser (Chameleon Vision S, Coherent) with a central wavelength of 800 nm. First, the laser beam is collimated by lenses L1 and L2 ($f_1$= 50 mm, $f_2$ = 50 mm). Next, a transmission grating G1 (T-1400-800, 1400 lines/mm, Lightsmyth) and a 4-f system (L3 and L4; $f_3$ = 100 mm, $f_4$ = 250 mm) are introduced to pre-compensate the angular dispersion induced by the DMD (DMD1, DLP 7000 0.7" XGA, Texas Instruments). After the DMD1 and lens L5 ($f_5$ = 200 mm), a spatial filter (SF) selects the -1[st] order diffraction beam in the Fourier plane, which contains the reconstructed wavefronts. Lastly, a 4-f system, i.e., lens L6 ($f_6$ = 54 mm) and the objective lens (OL1, EC Plan-NEOFLUAR 63×, NA = 1.25, Zeiss), relays the scanning laser to the liquid photoresin on a glass substrate, which is affixed to a precision XYZ stage. For the ODT module, a continuous wave (CW) laser beam with a central wavelength of 532 nm is used as the light source (MGL-III-532-300mW, CNI Lasers), which is coupled into a 1×2 single-mode fiber coupler (SMFC). The two output ends of the fiber coupler serve as the sample beam and reference beam, respectively. To ensure uniform sample illumination, the sample beam is collimated by a lens L7 ($f_7$ = 200 mm) and fully fills the aperture of the DMD (DMD2, DLP LightCrafter 9000, Texas Instruments). After DMD2, multiple diffraction orders are generated from the designed binary holograms displayed on DMD2. Next, lens L8 ($f_8$ = 150 mm) collects the diffraction orders and directs them to a third DMD (DMD3, DLP LightCrafter 6500, Texas Instruments), located on the Fourier plane of L8. Notably, here DMD3 functions as a dynamic spatial filter to select the 1[st] diffraction order (and reject redundant diffraction orders) to enter the subsequent optical system [24]. The filtered beam is reflected by a mirror (M1) and then collimated by a lens L9 ($f_9$= 200 mm). Lastly, lens L10 ($f_{10}$ = 300 mm) and the objective lens (OL2, LCI Plan-NEOFLUAR 63×, NA =

1.3, Zeiss) form a 4-f system to magnify the angle of the illumination beam over the sample to around 60 degrees. The scattered light from the sample is collected by OL1, which forms an intermediate image at the back focal plane of lens L11 ($f_{11}$ = 150 mm). In the end, the sample beam and the reference beam collimated by lens L12 ($f_{12}$ = 150 mm) are combined through a beam splitter (BS) and relayed to the camera through another 4-f system formed by lens L13 ($f_{13}$ = 60 mm) and L14 ($f_{14}$ = 400 mm). A high-speed camera (Fastcam SA-X2, Photron) is used to record the interferograms at a rate of 8,000 frames per second (fps). From the interferograms, complex sample fields corresponding to different illumination angles are retrieved. Lastly, by using an inverse scattering model based on the first-order Rytov approximation [15], a 3D RI map of the printed structure can be reconstructed. According to Abbe's diffraction limit, the lateral resolution of our system is estimated to be around 209 nm [28], while the axial resolution is approximately 2 times or more than the lateral resolution.

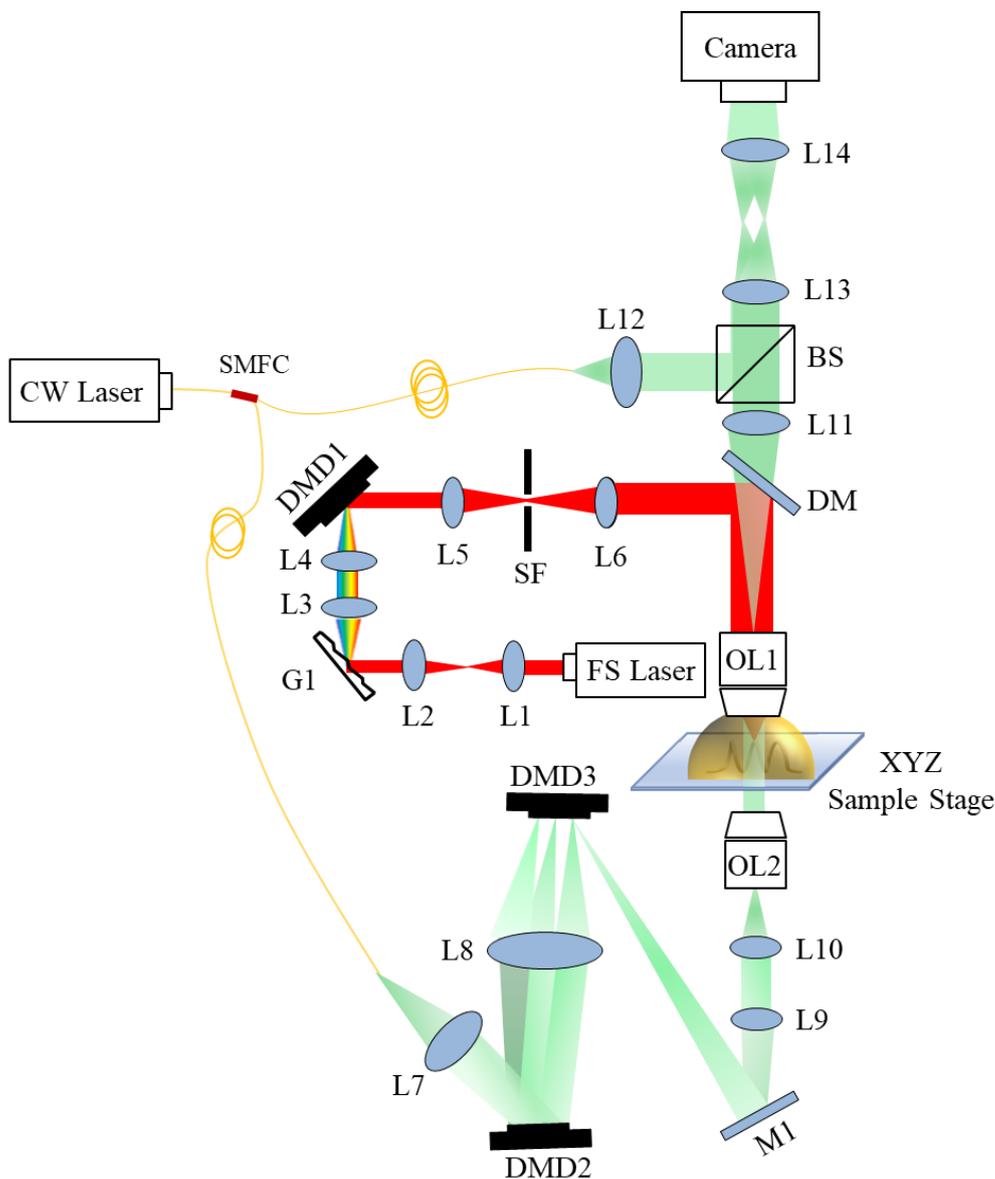

***Fig. 1.*** *Design of the integrated ODT-TPP system. L1-L14: lenses; G1: grating; DMD: digital micromirror device; SF: spatial filter; OL1-OL2: objective lens. SMFC: single-mode fiber coupler; M1: mirror; BS: beam splitter; and DM: dichroic mirror.*

## 3. 3D imaging validation

After the printing and imaging systems are optimized, we designed several 3D microstructures for subsequent experiments. To validate the 3D imaging capability of ODT, we firstly printed a standing along spiral which has a dimension of 49x17x17 μm$^3$ and a period of 11.7 μm, following the design in Fig. 2(a). The spiral was then imaged using bright-field microscopy and SEM as shown in Fig. 2(b) and (c), respectively. After that, we measured the 3D RI distribution with our ODT system. Figure 2(d) shows the 3D rendered RI map, from which the whole spiral structure can be clearly observed (refer to 3D rending and the image stack in Supplementary Video S1). Figure 2(e) shows an image stack of the RI map along the direction, denoted by the black arrow in Fig. 2(a). The structures revealed by ODT are consistent with the design, indicating the fabrication quality is high. Compared with bright-field microscopy and SEM, we revealed the entire 3D structure of the spiral.

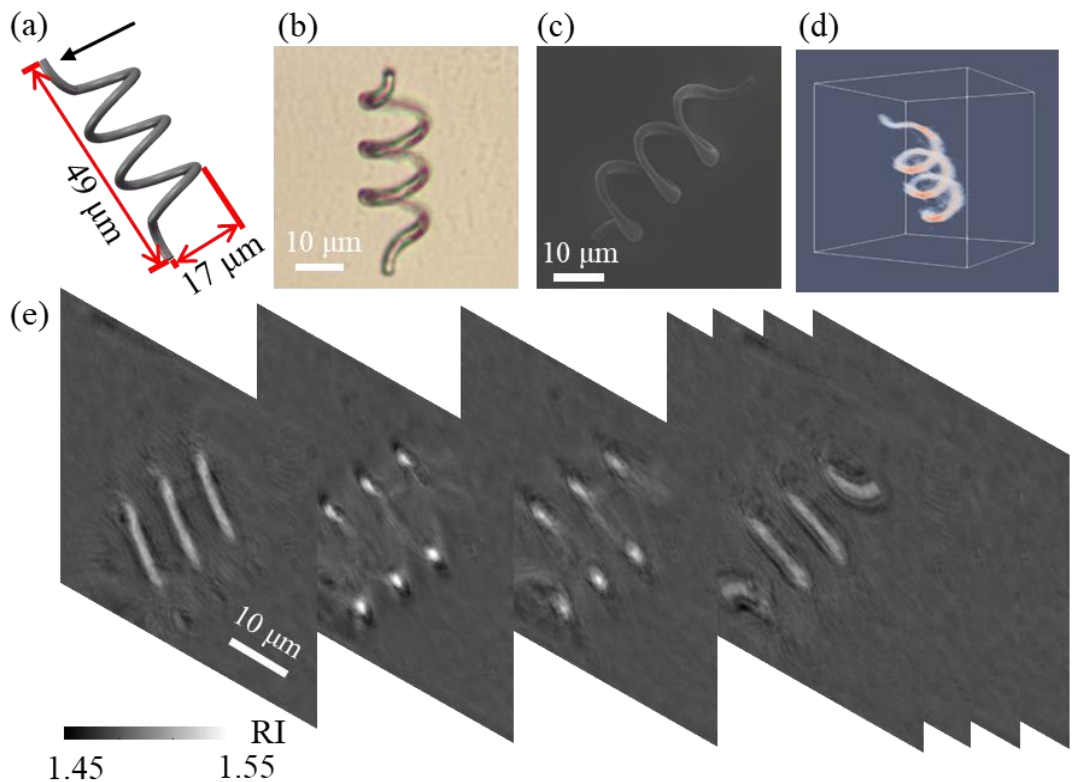

*Fig. 2.* Imaging results of a standing-along spiral. (a) Isometric view of the designed spiral. (b) Bright-field image of the printed spiral. (c) SEM image of the printed spiral. (d) A side view of the 3D rendered RI map of the printed spiral. (e) Image stack of the 3D RI map along the direction denoted by the black arrow in (a).

## 4. 3D imaging of enclosed spiral structure

After the validation, we printed a spiral and enclosed it in a shell, following the design in Fig. 3(a). The shell size is 20 x 18 x 16.5 μm$^3$ with a side wall thickness of 2.5 μm, a top wall thickness of 6 μm, and an open bottom that is attached to a No. 1 cover glass. The enclosed spiral has a dimension of 10.7 x 8 x 8 μm$^3$ and a period of 2.5 μm. Using our ODT system, the 3D morphology of the enclosed spiral in the shell is fully reconstructed. Figure 3(b) shows a top view of the 3D rendered RI map (refer to the 3D rendering and image stack video in Supplementary Video S2), while an image stack along the direction, indicated by the black arrow in Fig. 3(a), is shown in Fig. 3(c). From the image stack, the spiral structure is again clearly revealed. As a comparison, we also imaged this enclosed spiral under bright-field microscopy prior to conducting the ODT measurement. Figure 4(a) and (b) show the bright-field images when the sample is illuminated from the top and the bottom, respectively. As the reflected light by the spiral is weak, the spiral cannot be detected in Fig. 4(a). Although the spiral can be detected from the bottom (Fig. 4(b)), the image contrast is poor due to weak light absorption by the spiral structure. To prepare the sample for imaging with an SEM system, we performed sputter coating to make it conductive. Figure 4(c) and (d) show the top and side views from SEM, from which the 3D surfaces are clearly revealed, but the internal structures cannot be accessed at all. However, if the sample lacks a proper coating, the SEM images will appear blurring (Fig. 4(e) and (f)), especially for the side-view (Fig. 4 (f)). Note that we used the same sample when conducting SEM measurements by changing the coating. For this enclosed spiral, both bright-field and SEM failed to evaluate the fabrication quality of its features. This further demonstrates the advantages of ODT in fully revealing the 3D structures with good contrast without involving a complicated sample preparation process.

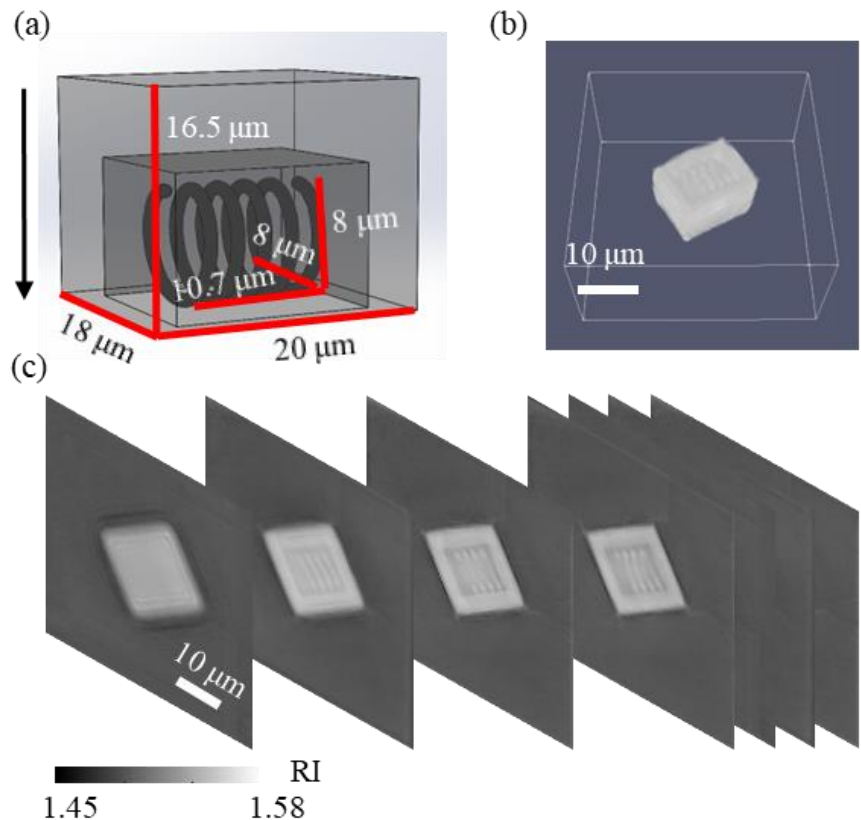

*Fig. 3. The ODT imaging results of the spiral enclosed in a shell. (a) Isometric view of the designed enclosed spiral. (b) Bird's eye view of the 3D rendered RI map of the printed enclosed spiral. (c) Image stack of the 3D RI map along the direction denoted by the black arrow in (a).*

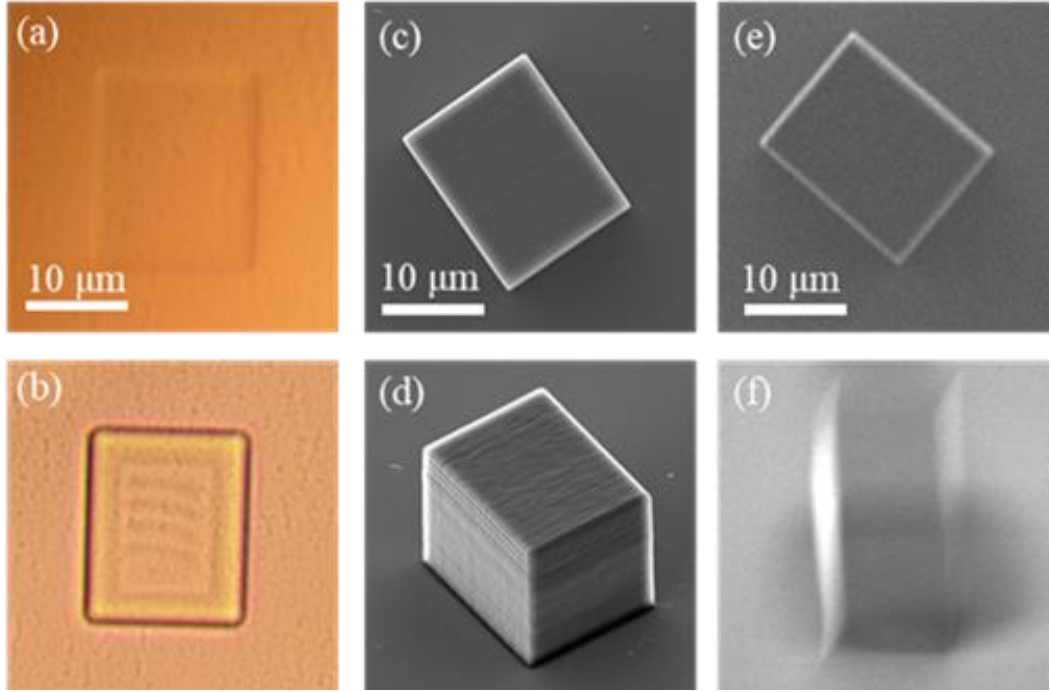

***Fig. 4.*** *The bright field light microscopy and SEM imaging results of the enclosed spiral in a shell. (a) and (b) are bright field light microscopy images when light is illuminated from the top and the bottom of the sample, respectively. (c) and (d) are top and side views of the spiral with shell structure with proper conductive coating imaged by SEM. (e) and (f) are top and side views of the same structure without proper conductive coating imaged by SEM.*

## 5. Refractive index measurement

Knowing the RI distributions of materials fabricated by TPP is essential for designing novel optical devices for specific applications, such as gradient-index elements for focusing, compound lenses for chromatic dispersion controlling, and micro waveguides for photonic integrated circuits [29]. In addition, the RI values of polymerized structures can be affected by the laser intensity, exposure time, and printing speed [13]. Thus, measuring RI changes will provide insights for us to optimize the laser parameters during the printing of structures.

To test the capability of our system in mapping the RI distributions, we printed three rectangular cuboids with each made from a unique commercial photoresin material (from Nanoscribe GmbH), following the design in Fig. 5(a), i.e., left (IP-L, RI round 1.518 at wavelength of 532 nm [30]), middle (IP-Dip, RI around 1.553 at wavelength of 532 nm [13, 30], and right (IP-PDMS, RI around 1.45 at wavelength of 589 nm; note that RI data at 532 nm is not available and we do expect a large difference in RI between 532 nm and 589 nm). Figure 5(b) and (c) are the SEM images at differvisuaent views. Figure 5(d)-(f) are the corresponding imaging results obtained from our ODT system, from which not only the 3D structural features are obtained, but also the RI values are mapped at all positions. Figure 5(d) is a bird's eye view of the 3D rendered RI map (refer to the 3D rendering in Supplementary Video S3). The x-y and x-z cross-sections, cut along the yellow and white dotted lines as pointed by the red arrows in Fig. 5(a), are shown in Fig. 5(e) and (f), respectively. After segmentation, we calculated the mean RI values of the IP-L, IP-Dip, and IP-PDMS regions, which are around 1.516, 1.531, and 1.474, respectively. The RI values are consistent with the reported values in Ref. [13, 30], and the small deviations might be due to the fact that the RI of the photoresin may vary with the actual curing conditions [13].

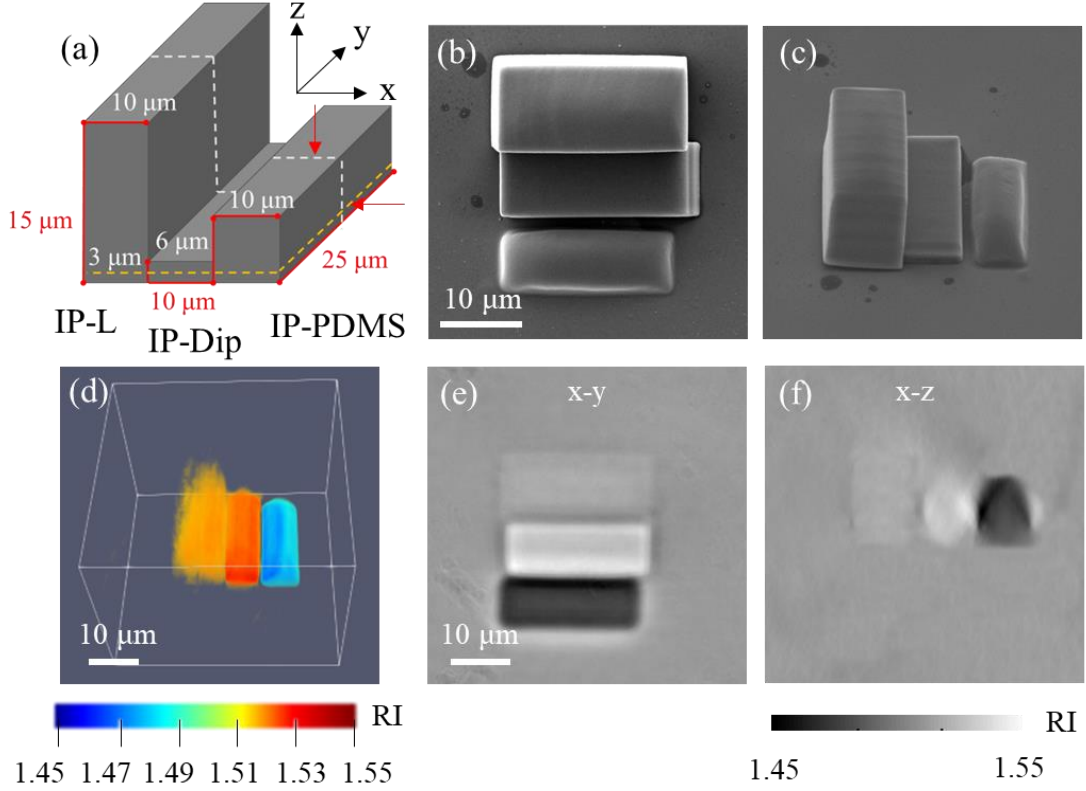

*Fig. 5. Refractive index measurements of the rectangular cuboids printed with different photoresins. (a) Isometric view of the designed rectangular cuboids. (b)&(c) Top and side views of the structure imaged by SEM. (d) A bird's eye view of the 3D rendered RI map measured from the structure. (e)-(f) Cross-sections of the reconstructed 3D RI map in (e) x–y plane (cut along the yellow dotted line in (a)) and (f) x-z plane (cut along the white dotted line in (a)).*

## 6. Surface roughness measurement

Surface roughness is a critical parameter for evaluating the performance of the printed structures for many optical applications. For example, the surface roughness of micro-lenses needs to be several times smaller than the wavelength of the testing light [31]. Several methods have been proposed for improving the surface roughness of printed structures, such as adaptive stitching [32] and trajectory optimization [33]. However, methods for characterizing the surface roughness of 3D printed structures are still lacking, especially for TPP printing.

6.1. Top surface roughness characterization

We printed a cube that has a dimension of 25 x 25 x 7 µm$^3$, following the design shown in Fig. 6(a), and measured its 3D RI map with the ODT system. Using the RI map, we reconstructed the surface profile as shown in Fig. 6(b), while a line profile along the red dotted line is shown in Fig. 6(c). After segmenting the cube height map from the background, we quantified its top surface roughness by calculating the root mean square error (RMSE), which is found to be around 81.7 nm. To validate the roughness measurement, we then used the AFM to map the surface profile over the same field of view (FOV) as the ODT, i.e., 56 x 56 µm$^2$, as shown in Fig. 6(d) - (f). Note that a default low-pass filter is applied in the AFM measurement software, so the curve appears smooth. The surface roughness obtained from AFM is around 81.4 nm, which is consistent with the results from our ODT method.

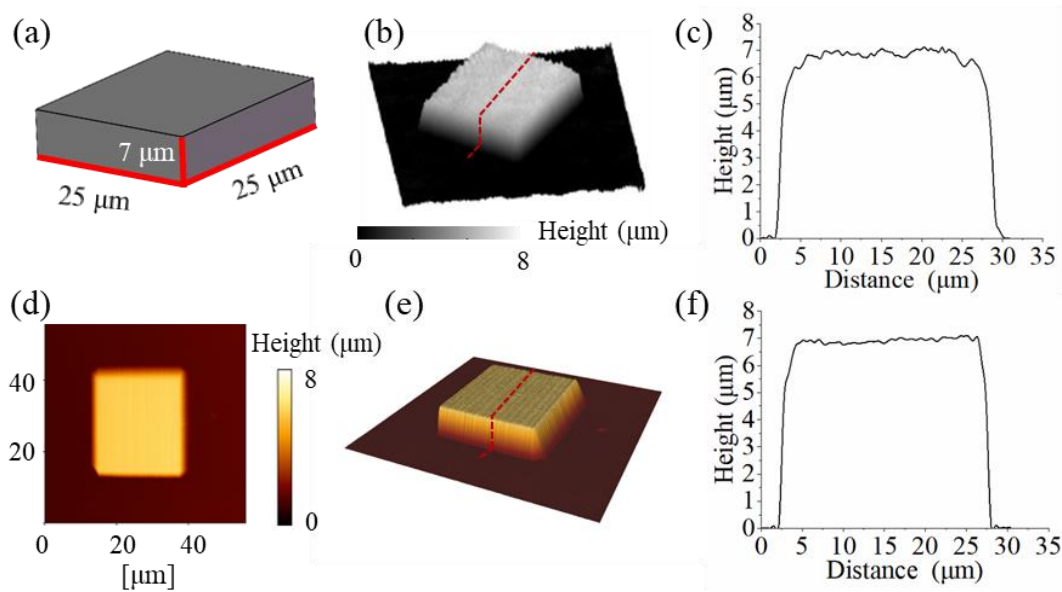

***Fig. 6.*** *Top surface roughness measurement of a cube using ODT. (a) Isometric view of the cube design. (b) The height map measured by ODT. (c) Line profile along the red dotting line in (b). (d)&(e) Top view and bird's eye view of the height map measured by AFM. (f) Line profile along the red dotted line in (e).*

6.2. Side wall's tilt and roughness characterization

AFM can well characterize the roughness of the sample top surfaces, but it cannot easily access the side walls. Here we demonstrate characterizing the side-wall roughness using ODT by designing a stacked tower structure as illustrated in Fig. 7(a)-(i). The tower consists of layers of different dimensions with a bottom layer dimension of 25 x 25 x 1 µm$^3$. As the layer goes up, the side length gradually decreases by 1 µm over each layer (i.e., 0.5 µm narrower than the layer below on both sides). When the side length reaches 15 µm, the side length then gradually increases by 1 µm over each layer until reaches back to 25 µm. Figure 7(b) shows the top view and two bird's eye views of the structure obtained by SEM. Figure 7(c) shows the measured 3D RI maps but visualized with similar perspectives as the SEM images. An image stack of the RI map, along the black arrow direction denoted in Fig. 7(a)-(i), is shown in Fig. 7 (d). It was found from the SEM images that the tower has a slight tilt of 0.25 µm from the original design, as indicated by the red arrow in Fig. 7(b)-(i). From the SEM images in Fig. 7(b)-(ii)&(iii), we further found blurred steps in the lower part of the tower, while clear steps were seen in the upper part of the tower. The deviation of the printed tower from the design is probably due to the slight tilt of the substrate, thus resulting in a dislocation between layers in one direction. The tilt and blurring of the lower part are also observed in the image stack in Figure 7 (d), as indicated by the red arrows (the upper part appears firstly, while the lower part exits last) and white rectangular boxes (lower part has less RI variations). From the zoom-in regions enclosed by red boxes in Fig. 7(d), we can also see the blurring in the lower part of the tower. Based on the SEM images, we assume the dislocation of the layers increases linearly along the +z dimension, so we can estimate the side-wall roughness of the top part of the tower that includes 11 layers (enclosed in the red rectangular box). Based on the tilt estimation of 0.25 μm from Fig. 7(b)-(i), the top layer edge difference with respect to the center layer is estimated to be (5+0.25/2) =5.125 μm (i.e., $h_{11}$; the originally designed $h_{11}$ is 5 μm). Then, the edge difference $h_i$ for the $i^{th}$ layer is $(i − 1) * 5.125/(N − 1)$, where N=11 is the total number of layers of the structure. The actual designed roughness of the structure side wall, characterized using the root-mean-square error (RMSE), is finally estimated to be about 1.6 μm. Note that RMSE is calculated against a reference design structure of the same tilt and dimension but with a smooth side wall as illustrated in Fig. 7(a)-(iii). From the segmented RI map, we calculated the RMSE of the upper part of the side wall, which is found to be about 1.6 μm that is consistent with our estimation.

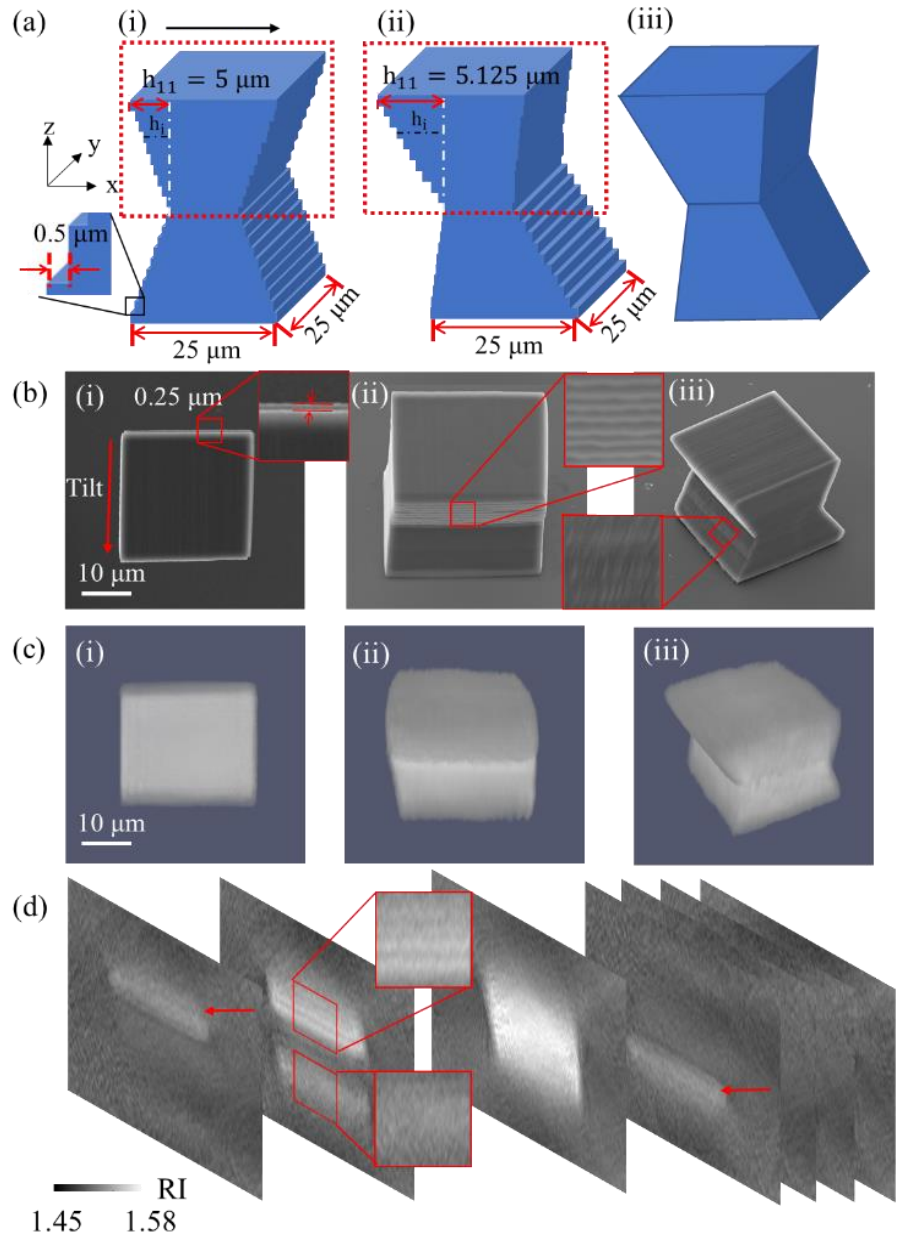

*Fig. 7.* Side wall roughness measurement of a stacked tower using ODT. (a): (i) Isometric view of the cube tower design; (ii) Isometric view of the inferred actual cube tower design with a tilt and blurring steps; and (iii) Isometric view of a reference design structure with a smooth side wall. (b): (i) Top view; (ii) bird's eye view; and (iii) isometric view of SEM images. (c): (i) Top view; (ii) bird's eye view; and (iii) isometric view of the 3D RI map measured by ODT. (d) Image stack of the 3D RI map along the black arrow direction (a)-(i).

## 7. Conclusion

In conclusion, we have presented a combined ODT-TPP system that can characterize both the external and internal structures of specifically designed 3D objects through mapping the 3D RI distributions with diffraction-limited resolution. In comparison, conventional characterization methods (e.g., SEM and AFM) have limited measuring capabilities, thus each method or even combinations of these methods cannot fully characterize the feature dimensions, internal structures, and surface roughness of the printed structures. Notably, the ODT measurement is performed in milliseconds, which is more than three orders of magnitude faster than conventional methods. Although the 3D image reconstruction and segmentation time are current around several minutes, it can be further expedited with parallel computing or machine learning algorithms. We envision our method can be potentially applied for large-scale characterization of micro- and nanostructures during the fabrication process and provide feedback for the optimization of the printing parameters in real-time.

## Acknowledgments

The authors acknowledge Croucher Foundation (Grant No. CM/CT/CF/CIA/0688/19ay); Hong Kong Innovation and Technology Fund (ITS/178/20FP & ITS/148/20); Hong Kong General Research Fund (No. 14209521 & 14209421); and Science, Technology and Innovation Commission of Shenzhen Municipality (STIC) (No. SGDX20201103095001009).

## CRediT authorship contribution statement

**Y. He:** Data curation, Formal analysis, Visualization, Writing - original draft, Writing - review & editing. **Q. Shao:** Data curation, Formal analysis, Writing - original draft, Writing - review & editing. **S-C. Chen:** Conceptualization, Funding acquisition, **R. Zhou:** Conceptualization, Funding acquisition, Supervision, Writing - review & editing.

## Declaration of Competing Interest

The authors declare that they have no known competing financial interests or personal relationships that could have appeared to influence the work reported in this paper.

## Funding

This work was supported by Croucher Foundation (Grant No. CM/CT/CF/CIA/0688/19ay); Hong Kong Innovation and Technology Fund (ITS/178/20FP & ITS/148/20); Hong Kong General Research Fund (No. 14209521 & 14209421); and Science, Technology and Innovation Commission of Shenzhen Municipality (STIC) (No. SGDX20201103095001009).

# Appendix A. Supplementary materials

**Video S1.** 3D rending and the image stack of a standing-along spiral

**Video S2.** 3D rending and the image stack of the spiral enclosed in a shell

**Video S3.** 3D rending of the rectangular cuboids printed with different photoresins